\documentclass[12pt]{iopart}
\usepackage{graphicx}
\usepackage{iopams}
\usepackage{psfrag}
\newcommand{\tj}{$t$--$J$}
\newcommand{\CuOtwo}{CuO$_2$}
\newcommand{\pair}{\tau}
\newcommand{\p}{^{\prime}}
\newcommand{\+}{^{\dagger}}
\newcommand{\pd}{^{\phantom{\dagger}}}
\newcommand{\kvec}{\bi{k}}
\newcommand{\rvec}{\bi{r}}
\newcommand{\Rvec}{\bi{R}}
\newcommand{\Qvec}{\bi{Q}}
\newcommand{\nullvec}{0}
\newcommand{\fksigmastate}{\vert \mbox{F}_{\kvec \sigma}\rangle}
\newcommand{\lksigmastate}{\vert \mbox{L}_{\kvec \sigma}\rangle}
\newcommand{\highoneholeenergy}{\epsilon^{(\mbox{\scriptsize high},1)}}
\newcommand{\hightwoholeenergy}{\epsilon^{(\mbox{\scriptsize high},2)}}
\newcommand{\lowoneholeenergy}{\epsilon^{(\mbox{\scriptsize low},1)}}

\newcommand{\nvec}{\bi{n}}
\newcommand{\ket}[1]{\vert {#1} \rangle}
\newcommand{\roottwo}{\surd 2}
\newcommand{\lambdac}{\lambda_{c}}
\newcommand{\boundstateenergy}{\epsilon}
\newcommand{\Ham}{\mathcal{H}}
\newcommand{\Greens}{\mathcal{G}}
\newcommand{\Gtilde}{\widetilde{\Greens}}
\newcommand{\Gprime}{\Greens\p}
\begin{document}
\title[Pairing from a pure Hubbard model]{A pure Hubbard model with demonstrable pairing adjacent to the Mott-insulating phase}
\author{J D Champion and M W Long}
\address{Theoretical Physics, School of Physics and Astronomy, The University of Birmingham,
         Edgbaston, Birmingham B15~2TT, UK}
\ead{champion@th.ph.bham.ac.uk}
\begin{abstract}
We introduce a Hubbard model on a particular class of geometries,
and consider the effect of doping the highly spin-degenerate Mott-insulating state with a microscopic
number of holes in the extreme strong-coupling limit.
The geometry is quite general, with pairs of atomic sites at each superlattice vertex, and a highly frustrated
inter-atomic connectivity: the one dimensional realization is a chain of edge-sharing tetrahedra.
The sole model parameter is the ratio of intra-pair to inter-pair hopping matrix elements.
If the intra-pair hopping is negligible then introducing a microscopic number of holes results in a ferromagnetic
Nagaoka groundstate.
Conversely, if the intra-pair hopping is comparable with the inter-pair hopping then the groundstate is
low spin with short-ranged spin correlations.
We exactly solve the correlated motion of a pair of holes in such a state and find that, in 1-d and 2-d,
they form a bound pair on a length scale that increases with diminishing binding energy.
This result is pertinent to the long-standing problem of hole-motion in the \CuOtwo\ planes of the high-temperature
superconductors: we have rigorously shown that, on our frustrated geometry, the holes pair up and a short-ranged
low-spin state is generated by hole motion alone.
\end{abstract}
\submitto{\JPA}
\section{Introduction}
In the years since the discovery of high-temperature superconductors, a wide spectrum of ideas
about the fundamental cause of superconductivity in \CuOtwo\ planes has been proposed,
but as yet none have been accepted as correct.
In this article we revisit Anderson's early suggestion of a resonating-valence-bond (RVB) state,
with `pre-formed' pairs~\cite{anderson87}, which at the time lacked rigorous support.

A suggestion currently popular in the literature is that the three-band Hubbard model in the strong-coupling
regime can be mapped onto the \tj\ model via the idea of Zhang-Rice singlets~\cite{zhangrice88}.
We believe that this reduction eliminates important physics: for example, the $t$-term alone can only
generate Nagaoka ferromagnetism~\cite{nagaoka66} --- in order to account for the observed low-spin state one has to rely
on the higher-order $J$ term.
In the accompanying article by IBS and MWL~\cite{iain03}, evidence is provided to support the hypothesis that
the three-band model allows an RVB state to be generated solely by the hole motion.

The three-band model for \CuOtwo\ planes is highly frustrated in a subtle way~\cite{long92}, and it is
this frustration that is lost in the reduction to the \tj\ model.
In this article we provide rigorous evidence for energetically favourable hole-pairing in a not dissimilar
highly-frustrated Hubbard model in the strong-coupling limit. The frustration is crucial, as it is this that
stabilizes the short-range correlated low-spin state, which is generated by hole motion alone.
%
%
\section{The model and geometry}

\begin{figure}
\begin{center}
\includegraphics[width=0.9\textwidth]{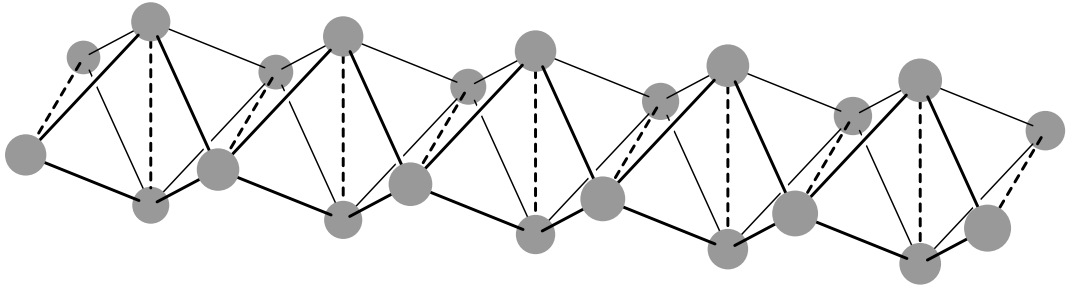}
\end{center}
\caption{\label{fig:tetrahedra}A chain of edge-sharing tetrahedra, a one-dimensional example of the
class of geometry considered in this article. The inter-pair hopping (\full) is, in general,
not the same strength as the intra-pair hopping (\dashed).}
\end{figure}

In this study we consider a repulsive Hubbard model on a particular class of highly-connected lattice.
The geometry is that of a superlattice of vertex pairs. In one dimension we will consider a linear-chain
superlattice, in two a square superlattice and in three a cubic superlattice.
The linear chain situation, where each superlattice site has a co-ordination number $Z=2$, can be realized
physically as a chain of edge-sharing tetrahedra, as illustrated in figure \ref{fig:tetrahedra}.

The single-particle part of the model Hamiltonian can be specified by the options open
to a charged spin-half fermion on any particular vertex.
The fermion can hop to any of the vertices in the $Z$ neighbouring pairs with a matrix element $-t$.
It can also hop to the other vertex in the pair with a matrix element which, for convenience,
we define to be $-Z\lambda t$.
The parameter $\lambda$  expresses the difference in energy scales between inter- and intra-pair hopping.
The electrostatic correlations between the charged fermions are included via an on-site Hubbard term:
two particles occupying the same vertex incur a Coulomb penalty $U$.
The second-quantized Hamiltonian is
\begin{equation}\fl
H = 
-t \sum_{\langle i,i\p \rangle} \sum_{\pair,\pair\p,\sigma} 
  p\+_{i\pair\sigma}\,p\pd_{i\p\pair\p\sigma}
-Z\lambda t \sum_{i,\pair,\sigma}
  p\+_{i\pair\sigma}p\pd_{i\bar{\pair}\sigma}
+\frac{U}{2} \sum_{i,\pair,\sigma} 
  p\+_{i\pair\sigma} p\pd_{i\pair\sigma} p\+_{i\pair\bar{\sigma}} p\pd_{i\pair\bar{\sigma}} \, ,
\end{equation}
where $p\+_{i\pair\sigma}$ creates a fermion of spin $\sigma$ on vertex $\pair$ of the pair on superlattice site $i$.
The complementary spin, or vertex, is denoted by a bar, and $\langle i,i\p \rangle$ indicates summation over nearest
neighbours $i$ and $i\p$.

We are concerned with behaviour adjacent to the Mott-insulator, and so it is appropriate to rewrite the Hamiltonian
in terms of fermionic hole operators, via the mapping $p\+_{i\pair\sigma} \mapsto h\pd_{i\pair\sigma} $.
Then, up to an ignorable constant, the hole Hamiltonian is
\begin{equation}\fl
H = 
t \sum_{\langle i,i\p \rangle} \sum_{\pair,\pair\p,\sigma} 
  h\+_{i\pair\sigma}\,h\pd_{i\p\pair\p\sigma}
+Z\lambda t \sum_{i,\pair,\sigma}
  h\+_{i\pair\sigma}h\pd_{i\bar{\pair}\sigma}
+\frac{U}{2} \sum_{i,\pair,\sigma} 
  h\+_{i\pair\sigma} h\pd_{i\pair\sigma} h\+_{i\pair\bar{\sigma}} h\pd_{i\pair\bar{\sigma}} \, .
\label{eq:holeHam}
\end{equation}

In this article we restrict attention to the extreme strong-coupling limit of $U=\infty$ with a finite positive $t$.
At maximal filling in the finite energy subspace the system is a Mott insulator with one particle
(or, equivalently, one hole) per vertex:
this state is hugely spin-degenerate because the super-exchange energy scale, $J$, is rigorously equal to zero.
We address the problem of how this degeneracy is lifted by the inclusion of a microscopic number of holes.
This particular geometry is useful because we can treat the doping of one and two holes analytically.
Before the more interesting issue of a two-hole bound state we will introduce the two competing styles of groundstate
via the more straightforward single-hole situation.
%
%
\section{Local symmetries and one-hole behaviour}
Eigenstates respect the symmetries of the Hamiltonian, and this is useful when categorizing
the possible styles of solution.
For example, the model that we consider in this article preserves the spin of a hole when it hops: as a result
of this global symmetry, the total spin of a state is conserved.
There also exists a collection of $N$ local symmetries, where $N$ is the number of superlattice sites:
these arise from the fact that the Hamiltonian is unchanged when the two vertices of any pair are swapped.
In looking for the groundstate of our Hamiltonian we categorize the eigenstates by their symmetry with respect
to these $N$ Hamiltonian-preserving operations.
We consider states that are either symmetric or antisymmetric with respect to all $N$ of the local symmetries:
we do not deal with the mixed-symmetry states.
\subsection{The high-spin state}
Setting the parameter $\lambda$ to zero effectively removes all the intra-pair bonds;
if the superlattice is bi-partite then the connectivity is no longer frustrated,
and hence the one-hole groundstate will exhibit Nagaoka ferromagnetism.

In the real-space representation that we choose to work with, every site is occupied with either a spinless
fermionic hole or a hard-core bosonic spin.
To illustrate this, consider a small part of the geometry comprising two nearest-neighbour vertex pairs
in the ferromagnetic one-hole state:
if the hole is on one of the vertices then the other three vertices accommodate parallel spins.
In pictorial form, the action of the Hamiltonian on this unit is simply
\begin{equation}\label{eq:hamoneholeferro}
\psfrag{h}{h}
\psfrag{o}{$\sigma$}
H
\includegraphics[width=1.25cm,viewport=0 50 130 130]{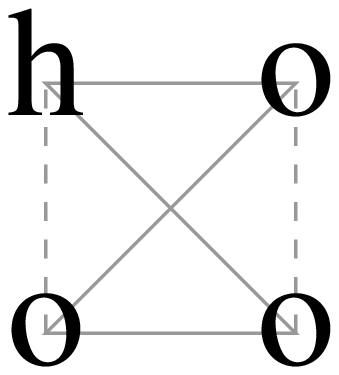} 
=
Z\lambda t
\includegraphics[width=1.25cm,viewport=0 50 130 130]{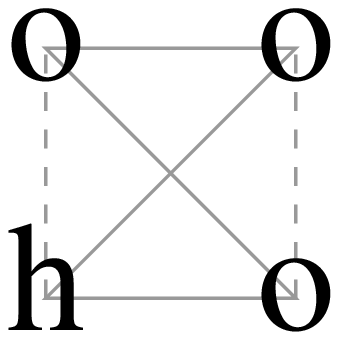} 
+ t \left [
\includegraphics[width=1.25cm,viewport=0 50 130 130]{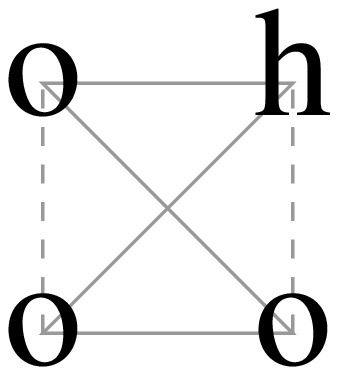} 
+ 
\includegraphics[width=1.25cm,viewport=0 50 130 130]{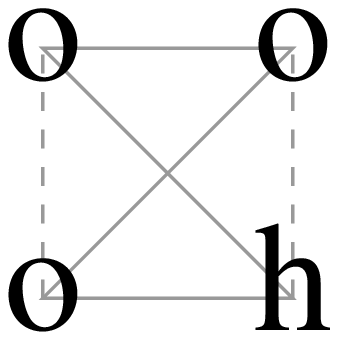} 
\right ] \, .
\end{equation}
We can use this representation to write down the Bloch state for one hole with
momentum $\kvec$ in a ferromagnetic background:
\begin{equation}\fl
\fksigmastate = 
\frac{1}{\sqrt{N}} \sum_i \rme^{\rmi \kvec.\rvec_i}\,
\frac{1}{\roottwo} \left [ 
\psfrag{h}{h}
\psfrag{o}{$\sigma$}
\psfrag{i}{$_i$}
\psfrag{i+1}{$_{i+1}$}
\psfrag{i-1}{$_{i-1}$}
\includegraphics[width=3.32cm,viewport=0 50 346 130]{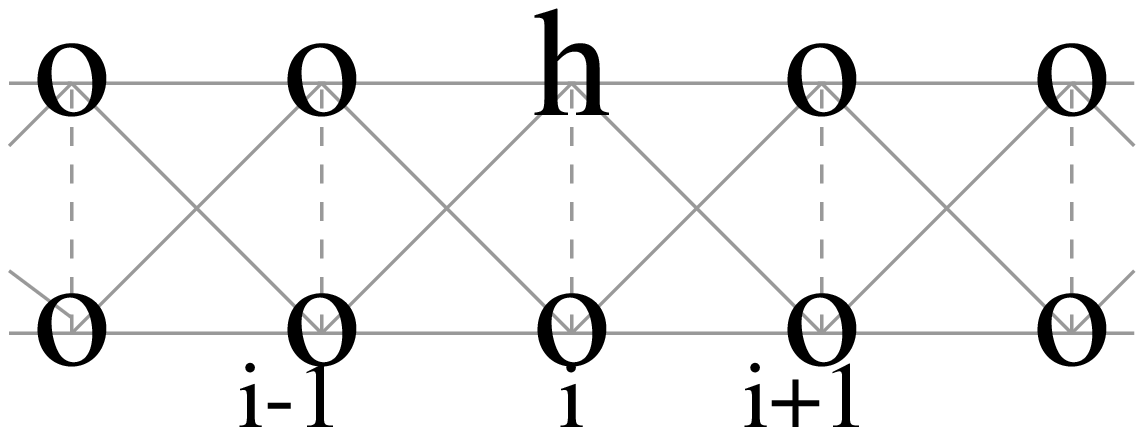} 
+
\includegraphics[width=3.32cm,viewport=0 50 346 130]{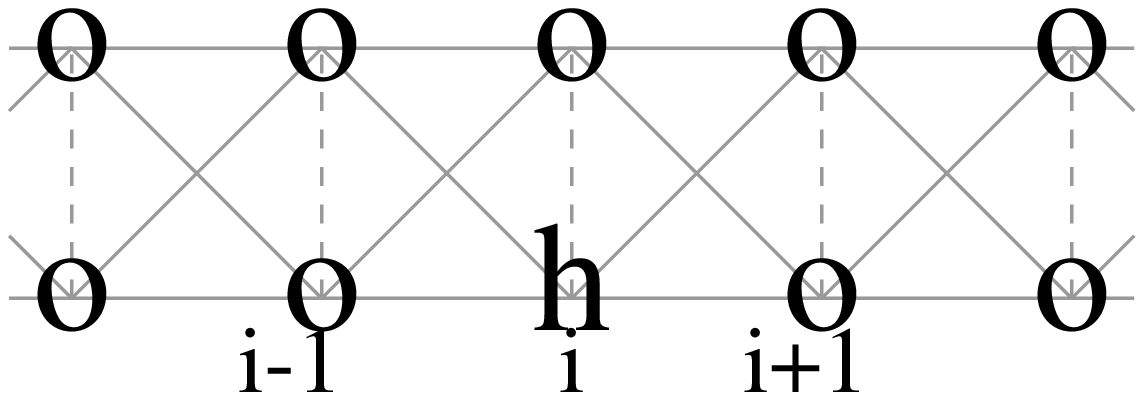} 
\right ]  .
\end{equation}
The one-dimensional lattice is used for illustration here because it is simple to depict;
it is obvious how this representation extends to the two- and three-dimensional lattices.
This state is `fully symmetric' since it maps onto itself when the vertices of any particular pair
are interchanged.

With the aid of equation (\ref{eq:hamoneholeferro}) it is straightforward to show that
$\fksigmastate$ is an eigenstate of the Hamiltonian with an energy eigenvalue
\begin{equation}
\highoneholeenergy_{\kvec\sigma} = \lambda  + 2 \gamma_{\kvec} \, .
\end{equation}
Throughout this article, energy eigenvalues, denoted by $\epsilon$, are `per hole' and in units of $Zt$.
The one-hole groundstate has the wavevector $\kvec$ that minimizes the superlattice
structure factor $\gamma_{\kvec}$:
for the cubic family of superlattices that we are considering here this means that
$\kvec=\Qvec\equiv(\pi,\pi,\ldots)$ and therefore
\begin{equation}\label{eq:highoneholeenergy}
\highoneholeenergy_{\Qvec\sigma} = \lambda - 2 \, .
\end{equation}
%
%
\subsection{The low spin state}\label{ssec:oneHoleLowSpin}
The other class of state that we consider is the complement of the high-spin state,
in that it has low-spin and is `fully anti-symmetric' with respect to the
vertex-pair-swapping symmetry.

In our chosen representation any real-space pair that is free from holes must accommodate
two bosonic spins: in order to make this pair of bosons antisymmetric, the spins must be correlated
in a total-spin singlet.
We depict the singlet in the following way:
\begin{equation}
\includegraphics[width=0.56cm,viewport=0 50 58 130]{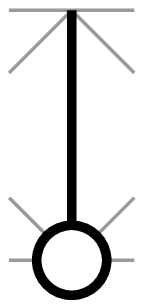}
\equiv 
\frac{1}{\roottwo}
\left[
{\psfrag{h}{$\uparrow$}\psfrag{k}{$\downarrow$}
\includegraphics[width=0.56cm,viewport=0 50 58 130]{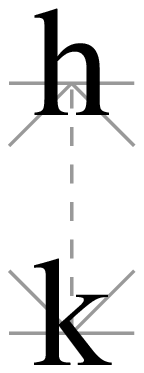} 
}
-
{\psfrag{k}{$\uparrow$}\psfrag{h}{$\downarrow$}
\includegraphics[width=0.56cm,viewport=0 50 58 130]{bosonic.eps} 
}
\right]
= (-1)
\includegraphics[width=0.56cm,viewport=0 50 58 130]{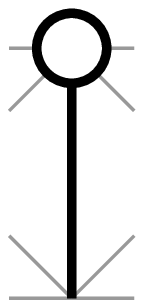}
\end{equation}
The antisymmetric property of the bosonic singlet is evident from the fact that swapping
the `head' and `tail' generates a factor of $-1$.

The one-hole antisymmetric low-spin state is
\begin{equation}\fl
\lksigmastate = 
\frac{1}{\sqrt{N}} \sum_i \rme^{\rmi \kvec.\rvec_i}\,
\frac{1}{\roottwo} \left [ 
\psfrag{h}{h}
\psfrag{o}{$\sigma$}
\psfrag{i}{$_i$}
\psfrag{i+1}{$_{i+1}$}
\psfrag{i-1}{$_{i-1}$}
\includegraphics[width=3.32cm,viewport=0 50 346 130]{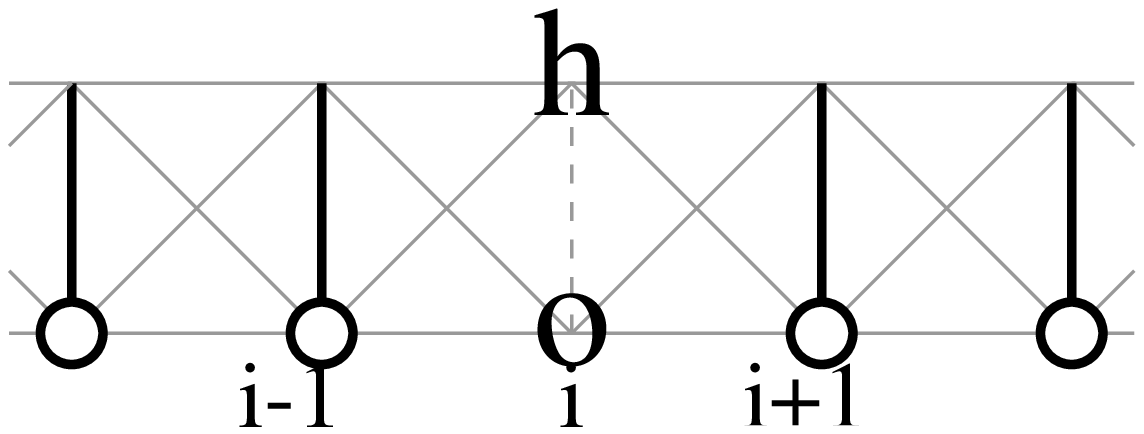} 
-
\includegraphics[width=3.32cm,viewport=0 50 346 130]{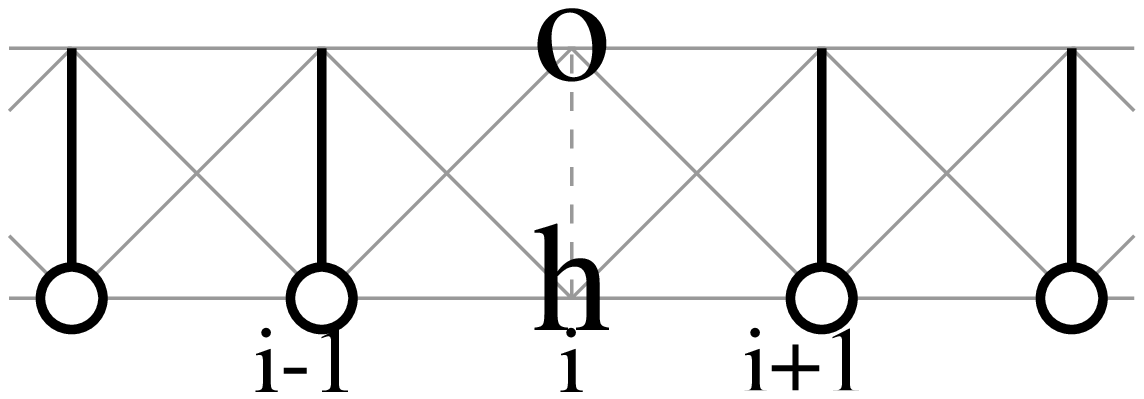} 
\right ]
\end{equation}
where the anti-phase combination guarantees antisymmetry with respect to vertex exchange on
the hole's pair.

The action of the Hamiltonian on a two-pair unit that contains the hole is
\begin{equation}
\psfrag{h}{h}
\psfrag{o}{$\sigma$}
H
\includegraphics[width=1.25cm,viewport=0 50 130 130]{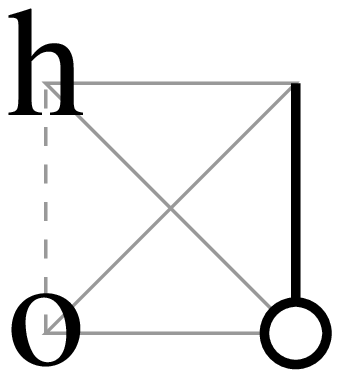} 
=
Z\lambda t
\includegraphics[width=1.25cm,viewport=0 50 130 130]{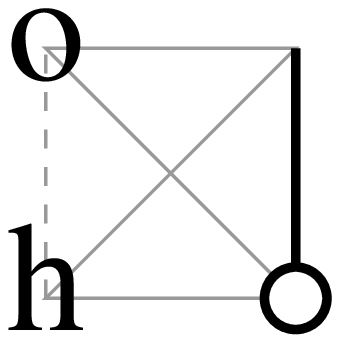} 
+ 
t
\left[
\includegraphics[width=1.25cm,viewport=0 50 130 130]{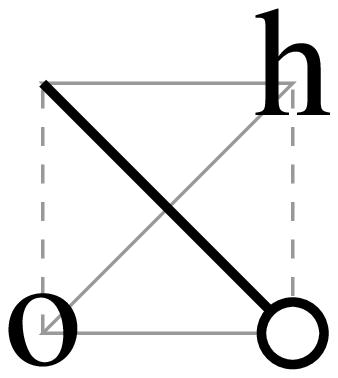} 
+
\includegraphics[width=1.25cm,viewport=0 50 130 130]{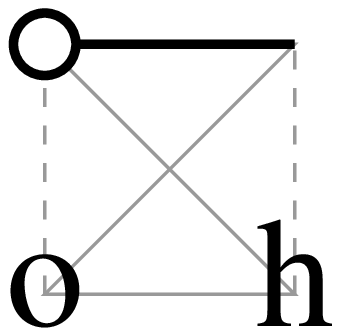} 
\right ]
\end{equation}
Acting the Hamiltonian on the state $\lksigmastate$ creates states where singlets
extend from one superlattice site to another.
Such states have a non-zero overlap with the original state, and we can use the
spin identity
\begin{equation}
\psfrag{h}{h}
\psfrag{o}{$\sigma$}
\includegraphics[width=1.25cm,viewport=0 50 130 130]{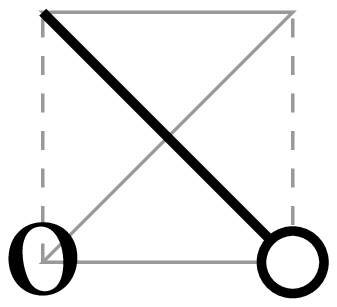}
-
\includegraphics[width=1.25cm,viewport=0 50 130 130]{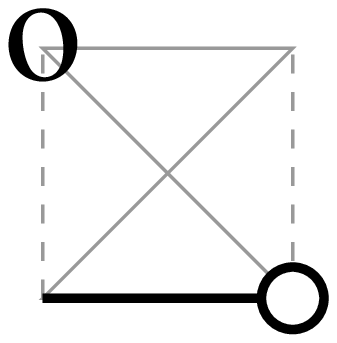}
=
\includegraphics[width=1.25cm,viewport=0 50 130 130]{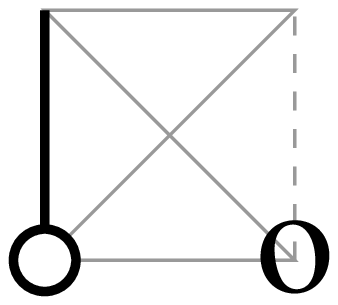}
\phantom{\int}
\end{equation}
to show that $\lksigmastate$ is in fact an eigenstate of the Hamiltonian with
energy eigenvalue
\begin{equation}\label{eq:lowoneholeenergy}
\lowoneholeenergy_{\kvec \sigma} = -\lambda + \gamma_{\kvec} .
\end{equation}
As was the case for the ferromagnetic one-hole state, the lowest energy state has $\kvec=\Qvec$.
By comparing equations~(\ref{eq:highoneholeenergy}) and (\ref{eq:lowoneholeenergy})
it is obvious that the groundstate is ferromagnetic when $\lambda < 1/2$ and low-spin with
short-range correlations when $\lambda > 1/2$.
%
%
\section{Two-hole behaviour}
In the previous section we saw that the one-hole solutions were trivial to find
once a suitable representation had been established. With this framework in place,
we are now ready to tackle the two-hole problem.
The high-spin state is simple to describe because the interaction is irrelevant and we
simply occupy a single non-interacting band;
the low-spin problem is not so straightforward, and we make use of the 
Greens function impurity technique to deal with the interaction.
\subsection{The high spin state}
The single hole problem in the symmetric subspace was solved by including a symmetric hole with
wavevector $\Qvec$; all other sites feature a spin $\sigma$ hence the state was described as `high spin'.
If we include another symmetric hole from the corner of the Brillouin zone, with the same spin as
the first, then Pauli exclusion prevents double-occupancy of any vertex.
On a macroscopic lattice these two holes, which are in orthogonal single-hole states, can have an
arbitrarily close energy. Hence the lowest energy \emph{per hole} of the two-hole high-spin state is
the same as the single-hole minimum, namely
\begin{equation}
\hightwoholeenergy_{\Qvec\sigma} = \lambda - 2 \, .
\end{equation}
%
%
\subsection{The low spin state}
Here we restrict attention to states where the pair of fermionic holes, h$_1$ and h$_2$,
have zero total momentum. The other degree of freedom is the separation between the two holes
on the superlattice, $\Rvec = \rvec_2 - \rvec_1$, and we use this quantity to label the low-spin
state basis.

Consider the $\Rvec = \nullvec$ state which, in its properly normalised form, is
\begin{equation}
\ket{\Rvec = \nullvec} = \frac{1}{\sqrt{N}} \sum_i 
\psfrag{h}{h$_1$}
\psfrag{k}{h$_2$}
\psfrag{i}{$_i$}
\psfrag{i+1}{$_{i+1}$}
\psfrag{i-1}{$_{i-1}$}
\includegraphics[width=3.32cm,viewport=0 50 346 130]{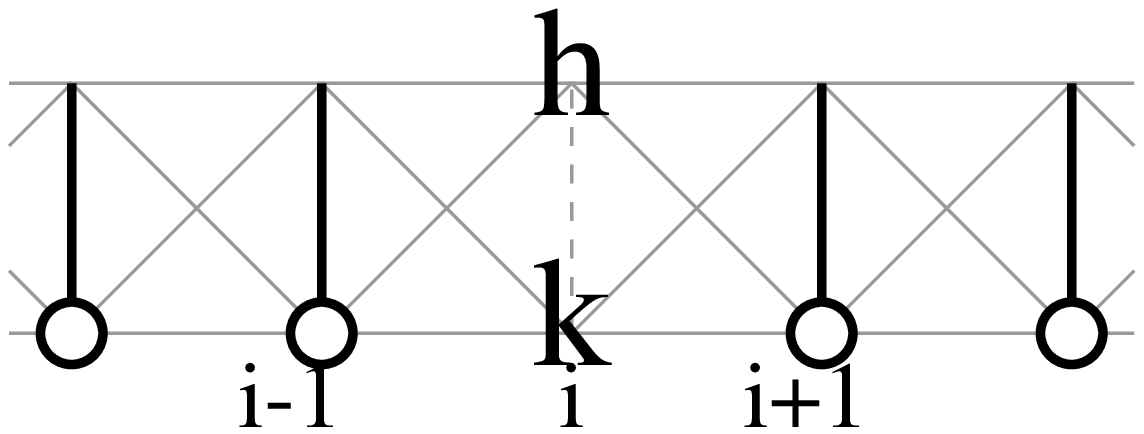} .
\end{equation}
This state has total-spin zero, short-range spin correlations, and is overall antisymmetric
since swapping a singlet's head and tail, or exchanging h$_1$ and h$_2$, result in a minus sign.

As before, consider a unit of this state comprising the vertex pair containing h$_1$ and h$_2$, and
one of its nearest neighbour pairs that must feature a singlet. The action of the Hamiltonian on
this unit can be written as
\begin{equation}
\psfrag{h}{h$_1$}
\psfrag{k}{h$_2$}
H
\includegraphics[width=1.25cm,viewport=0 50 130 130]{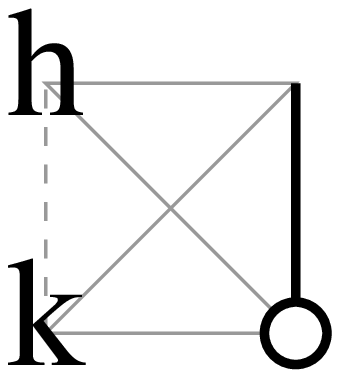} 
=
 t \left [
\includegraphics[width=1.25cm,viewport=0 50 130 130]{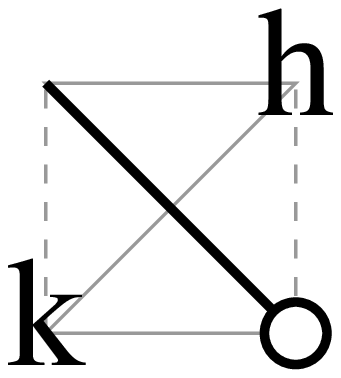} 
+ 
\includegraphics[width=1.25cm,viewport=0 50 130 130]{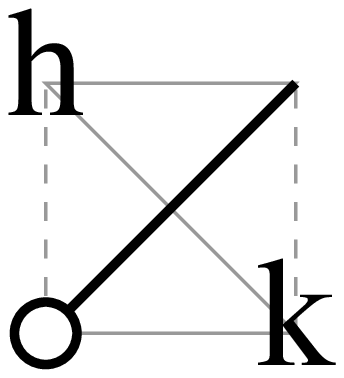} 
+
\includegraphics[width=1.25cm,viewport=0 50 130 130]{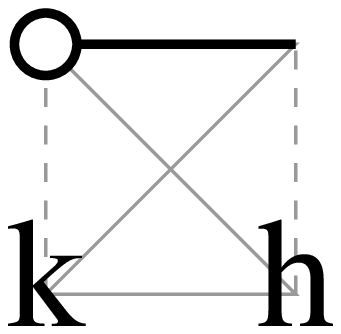} 
+
\includegraphics[width=1.25cm,viewport=0 50 130 130]{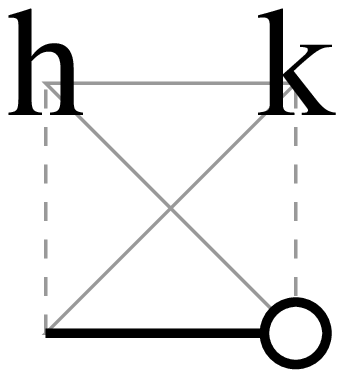} 
\right ] \, .
\end{equation}
Note that no use is made of the intra-pair hopping, as this is prohibited by
the $U=\infty$ Coulomb penalty.

Now, any particular superlattice site has $Z$ nearest neighbours; let us denote the
$Z$ vectors that translate to the nearest neighbours as
$\nvec_1, \nvec_2, \ldots, \nvec_Z$. We then choose to define the overall antisymmetric
two-hole state where the holes are separated by one lattice spacing as
\begin{equation}\fl
\ket{\Rvec \neq \nullvec} = \frac{1}{\sqrt{N}} \sum_i \frac{1}{2} \left[
\psfrag{h}{h$_1$}
\psfrag{k}{h$_2$}
\psfrag{i}{$_i$}
\psfrag{f}{$\uparrow$}
\psfrag{d}{$\downarrow$}
\includegraphics[width=1.25cm,viewport=0 50 130 130]{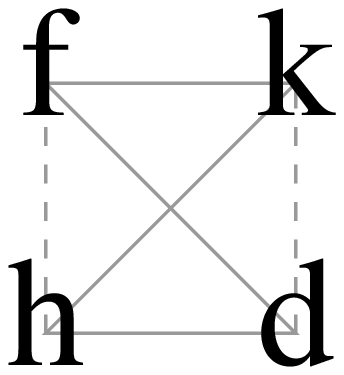} +
\includegraphics[width=1.25cm,viewport=0 50 130 130]{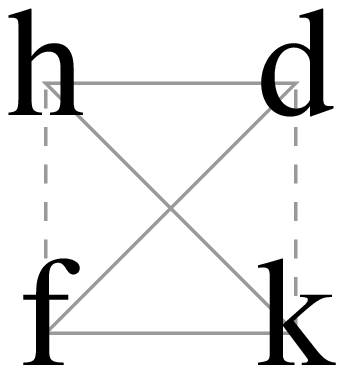} -
\includegraphics[width=1.25cm,viewport=0 50 130 130]{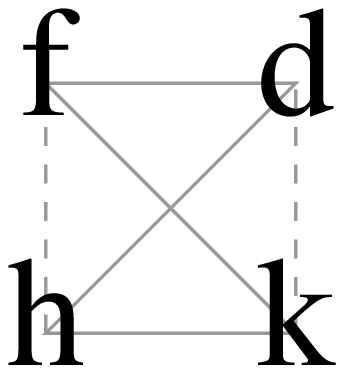} -
\includegraphics[width=1.25cm,viewport=0 50 130 130]{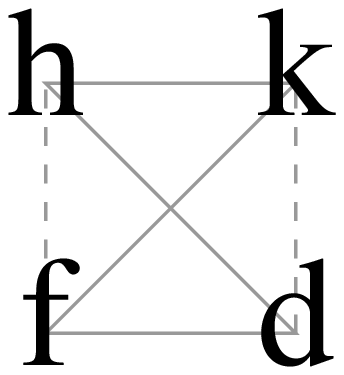} 
\right]
\end{equation}
where it is implied that all other vertex-pairs feature a spin singlet.
The convention is to put an $\uparrow$ spin on the same pair as the hole h$_1$,
and a $\downarrow$ with h$_2$; the factor of $1/2$ ensures correct normalisation.
It is then possible to express the outcome of applying the Hamiltonian to
$\ket{\Rvec=\nullvec}$ as
\begin{equation}
H \,\ket{\Rvec = \nullvec} = 2 \roottwo \, t \sum_{j=1}^{Z} \ket{\Rvec=\nvec_j} .
\end{equation}

The states for all other $\Rvec$, i.e.\ $\Rvec \neq \nullvec, \nvec_1, \nvec_2, \ldots, \nvec_Z$,
are defined in a similar way to $\ket{\Rvec=\nvec_j}$. The fact that these states
have the holes more than one hop apart means that they move as `independent holes'. The reasoning from
section \ref{ssec:oneHoleLowSpin} applies, and thus
\begin{equation}
H \,\ket{\Rvec \neq \nullvec, \nvec_1, \nvec_2, \ldots, \nvec_Z } =
-2Z\lambda\,t\,\ket{\Rvec} + 2\,t \sum_{j=1}^Z \ket{\Rvec+\nvec_j}
\end{equation}
where the factor of two arises because there are \emph{two} independent holes.

The action of the Hamiltonian on the nearest-neighbour states is a combination of the two above styles
of behaviour. The holes are on different vertex pairs, so they can gain the intra-pair hopping energy.
Hopping in one particular direction will bring both holes onto the same site, and hopping
in all the other directions is `independent-hole'-like. The result is that
\begin{equation}
H \,\ket{\Rvec = \nvec_p} = -2Z\lambda\,t\,\ket{\Rvec} + 2\roottwo\,t\,\ket{\nullvec}
                          +2\,t\sum_{j \neq p} \ket{\Rvec + \nvec_j} \, .
\end{equation}

Now that we know all the matrix elements, it can be seen that they deviate from a perfectly
periodic Hamiltonian in just two ways:
 (i) the coupling between $\ket{\nullvec}$ and the `nearest-neighbour' states $\ket{\nvec_j}$ is enhanced
     by a factor of $\roottwo$, and
 (ii) the intra-pair energy $2Z\lambda t$ is \emph{not} gained by $\ket{\nullvec}$.
We can write the Hamiltonian as the sum of a periodic term, $H_0$, and a `local' term $H_1$,
where the matrix elements of these two terms are given by
\begin{equation}
H_0 \ket{\Rvec} = -2Z\lambda\,t \, \ket{\Rvec} + 2\,t\sum_{j=1}^{Z} \ket{\Rvec + \nvec_j}
\end{equation}
and
\begin{equation}\fl
H_1 \left[
\begin{array}{c}
\ket{ \nullvec } \\ \ket{ \nvec_1 } \\ \ket{ \nvec_2 } \\ \vdots \\ \ket{ \nvec_Z }
\end{array}
\right]
= 2t
\left[
\begin{array}{ccccc}
Z\lambda   & \roottwo-1 &  \roottwo-1 & \cdots & \roottwo-1 \\
\roottwo-1 &  0         &      0      & \cdots &     0      \\
\roottwo-1 &  0         &      0      & \cdots &     0      \\
\vdots     &  \vdots    &   \vdots    & \ddots &  \vdots    \\
\roottwo-1 &  0         &      0      & \cdots &     0      
\end{array}
\right]
\left[
\begin{array}{c}
\ket{ \nullvec } \\ \ket{ \nvec_1 } \\ \ket{ \nvec_2 } \\ \vdots \\ \ket{ \nvec_Z }
\end{array}
\right] \, .\label{eq:H1matrix}
\end{equation}

It is convenient to rescale the Hamiltonian so that all energies are `per hole' in units of $Zt$:
the rescaled two-hole Hamiltonian, $\Ham$, is defined by $H = 2Zt\,\Ham$.
In order to exactly solve this eigenproblem we will make use of the Greens function for the entire
Hamiltonian, $\Ham = \Ham_0 + \Ham_1$, which can be written as
\begin{equation}
\Greens(z) = \Greens_0(z) +  \Greens_0(z) \, \Sigma(z) \, \Greens_0(z)
\end{equation}
where $\Greens_0(z)$ is the Greens function for $\Ham_0$, the periodic part of the Hamiltonian,
\begin{equation}
\Greens_0(z) = [z-\Ham_0]^{-1} = \frac{1}{z+\lambda - \gamma_{\kvec}}
\end{equation}
and
\begin{equation}
\Sigma(z) = [1 - \Ham_1 \Greens_0(z) ]^{-1} \, \Ham_1 .
\end{equation}
The energy eigenvalues of the periodic case alone are found at the poles of $\Greens_0(z)$.
If we anticipate that the inclusion of $\Ham_1$ generates a new state, then its pole must
be present in $\Sigma(z)$. Hence the pole is found at $z=\boundstateenergy$ such that
\begin{equation}
\det \, [ 1 - \Ham_1 \Greens_0(\boundstateenergy) ] = 0 \label{eq:greensdet} .
\end{equation}
Equation~(\ref{eq:H1matrix}) shows how the local term of the Hamiltonian, $\Ham_1$, can be represented
by a $(Z+1)\times(Z+1)$ matrix; the above determinant equation involves the product $\Ham_1 \Greens_0$, hence
the relevant part of the matrix $\Greens_0(z)$ is simply
\begin{equation}
\Greens_0(z) = \left[
\begin{array}{ccccc}
\Greens_{00}(z) & \Greens_{01}(z)  & \Greens_{02}(z) & \cdots & \Greens_{0Z}(z) \\
\Greens_{10}(z) & \Greens_{11}(z)  & \Greens_{12}(z) & \cdots & \Greens_{1Z}(z) \\
\Greens_{20}(z) & \Greens_{21}(z)  & \Greens_{22}(z) & \cdots & \Greens_{2Z}(z) \\
\vdots & \vdots  & \vdots & \ddots & \vdots \\
\Greens_{Z0}(z) & \Greens_{Z1}(z)  & \Greens_{Z2}(z) & \cdots & \Greens_{ZZ}(z)
\end{array}
\right]
\end{equation}
where, if we conveniently let $\nvec_0 = \nullvec$, these real-space elements are defined by
\begin{equation}
\Greens_{ij}(z) = \frac{1}{N} \sum_{\kvec} 
  \frac{\exp[\rmi\,\kvec.(\nvec_i-\nvec_j)]}{z+\lambda - \gamma_{\kvec}} .
\end{equation}
Extracting the point-symmetry of the superlattice reduces the matrices to 2$\times$2,
\begin{equation}
\Ham_1 = \left[
\begin{array}{cc}
\lambda & \roottwo -1 \\
\roottwo -1 & 0
\end{array}
\right]
\quad \mbox{ and } \quad
\Greens_0 = \left[
\begin{array}{cc}
\Greens_{00}(z) & \Gprime(z) \\
\Gprime(z)    & \Gtilde(z) 
\end{array}
\right]
\end{equation}
where
\begin{equation}
\Gprime(z) = \frac{1}{N} \sum_{\kvec} \frac{\gamma_{\kvec}}{z + \lambda - \gamma_{\kvec}} 
\quad \mbox{ and } \quad
\Gtilde(z) = \frac{1}{N} \sum_{\kvec} \frac{\gamma^2_{\kvec}}{z + \lambda - \gamma_{\kvec}} .
\end{equation}
Evaluating the determinant equation (\ref{eq:greensdet}) to find the pole at $z=\boundstateenergy$ gives
\begin{equation}
1 = \lambda \Greens_{00}(\boundstateenergy) + 2\,(\roottwo-1)\,\Gprime(\boundstateenergy) 
    + (\roottwo-1)^2\, [\Greens_{00}(\boundstateenergy)\, \Gtilde(\boundstateenergy) - {\Gprime}^2(\boundstateenergy)]
\end{equation}
which can be further simplified using the identities
\begin{equation}
\Gprime(\boundstateenergy) = 
  (\boundstateenergy+\lambda)\, \Greens_{00}(\boundstateenergy)-1
\quad \mbox{and} \quad 
  \Gtilde(\boundstateenergy)\, \Greens_{00}(\boundstateenergy) - {\Gprime}^2(\boundstateenergy) 
  = \Gprime(\boundstateenergy)
\end{equation}
to give the self consistent equation
\begin{equation}\label{eq:selfconsistent}
\frac{2}{\boundstateenergy + 2\lambda} = 
\frac{1}{N} \sum_{\kvec} \frac{1}{\boundstateenergy + \lambda - \gamma_{\kvec}} .
\end{equation}

\begin{figure}
\begin{center}
\psfrag{l1}[c][c]{\footnotesize{$\quad\lambdac^{(1)}$}}
\psfrag{l2}[c][c]{\footnotesize{$\quad\lambdac^{(2)}$}}
\psfrag{Energy / Zt}[c][c]{\footnotesize{Energy per hole}}
\psfrag{lambda}[c][c]{\footnotesize{$\lambda$}}
\psfrag{0}[c][c]{\footnotesize{$0$}}
\psfrag{0.2}[c][c]{\footnotesize{$0.2$}}
\psfrag{0.4}[c][c]{\footnotesize{$0.4$}}
\psfrag{0.6}[c][c]{\footnotesize{$0.6$}}
\psfrag{0.8}[c][c]{\footnotesize{$0.8$}}
\psfrag{1}[c][c]{\footnotesize{$1$}}
\psfrag{-2}[r][r]{\footnotesize{$-2$}}
\psfrag{-1.8}[r][r]{\footnotesize{$-1.8$}}
\psfrag{-1.6}[r][r]{\footnotesize{$-1.6$}}
\psfrag{-1.4}[r][r]{\footnotesize{$-1.4$}}
\psfrag{-1.2}[r][r]{\footnotesize{$-1.2$}}
\psfrag{-1}[r][r]{\footnotesize{$-1$}}
\includegraphics[width=0.75\textwidth,viewport=0 0 265 265]{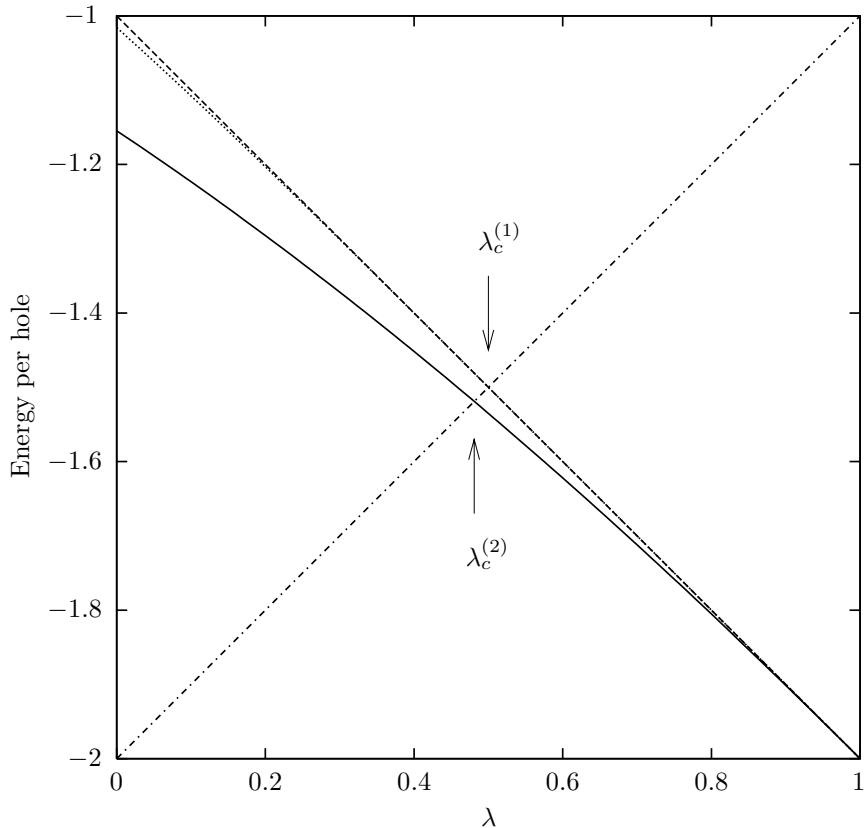}
\end{center}
\caption{\label{fig:onedbinding} The energy \emph{per hole} of various states as a function of the model
parameter $\lambda$.
Whether we are considering the linear chain, square, or cubic superlattice, the one-hole ferromagnetic Nagaoka
state (\chain) and one-hole low-spin state (\dashed) are degenerate at $\lambdac^{(1)}=1/2$.
The two-hole bound-pair low-spin state on the linear chain (\full) is degenerate with the two-hole Nagaoka state
at $\lambdac^{(2)}=0.4809(6)$.
The binding energy of the two-hole low-spin state on the square superlattice (\dotted) is an order of magnitude
smaller than the linear chain case;
in both 1-d and 2-d the binding energy vanishes at $\lambda=1$.}
\end{figure}

In one- and two-dimensions this equation has a solution in the parameter range $0 \leq \lambda < 1$.
For example, the energy per hole of the linear-chain superlattice state is given exactly by
\begin{equation}
\boundstateenergy = -\frac{2}{3} \left [ \lambda + \sqrt{\lambda^2 + 3} \right ]  
 \equiv -1 - \lambda-\delta \, ,
\end{equation}
where we have parameterized the binding energy per hole as $\delta$:
this energy is compared with the independent-hole low-spin and high-spin energies in 
figure~\ref{fig:onedbinding}. 
Equation~(\ref{eq:selfconsistent}) can be solved numerically for the case of the square superlattice;
the binding energy is about an order of magnitude smaller than the 1-d binding energy, and hence the
2-d bound state curve is barely visible below the independent-hole line in figure~\ref{fig:onedbinding}.

\begin{figure}
\begin{center}
\psfrag{lambda}[c][c]{\footnotesize{$\lambda$}}
\psfrag{xi}[c][c]{\footnotesize{$\xi$}}
\psfrag{0}[c][c]{\footnotesize{$0$}}
\psfrag{0.2}[c][c]{\footnotesize{$0.2$}}
\psfrag{0.4}[c][c]{\footnotesize{$0.4$}}
\psfrag{0.6}[c][c]{\footnotesize{$0.6$}}
\psfrag{0.8}[c][c]{\footnotesize{$0.8$}}
\psfrag{1}[c][c]{\footnotesize{$1$}}
\psfrag{5}[r][r]{\footnotesize{$5$}}
\psfrag{10}[r][r]{\footnotesize{$10$}}
\psfrag{15}[r][r]{\footnotesize{$15$}}
\psfrag{20}[r][r]{\footnotesize{$20$}}
\psfrag{25}[r][r]{\footnotesize{$25$}}
\psfrag{30}[r][r]{\footnotesize{$30$}}
\psfrag{35}[r][r]{\footnotesize{$35$}}
\psfrag{40}[r][r]{\footnotesize{$40$}}
\includegraphics[width=0.75\textwidth,viewport=0 0 265 265]{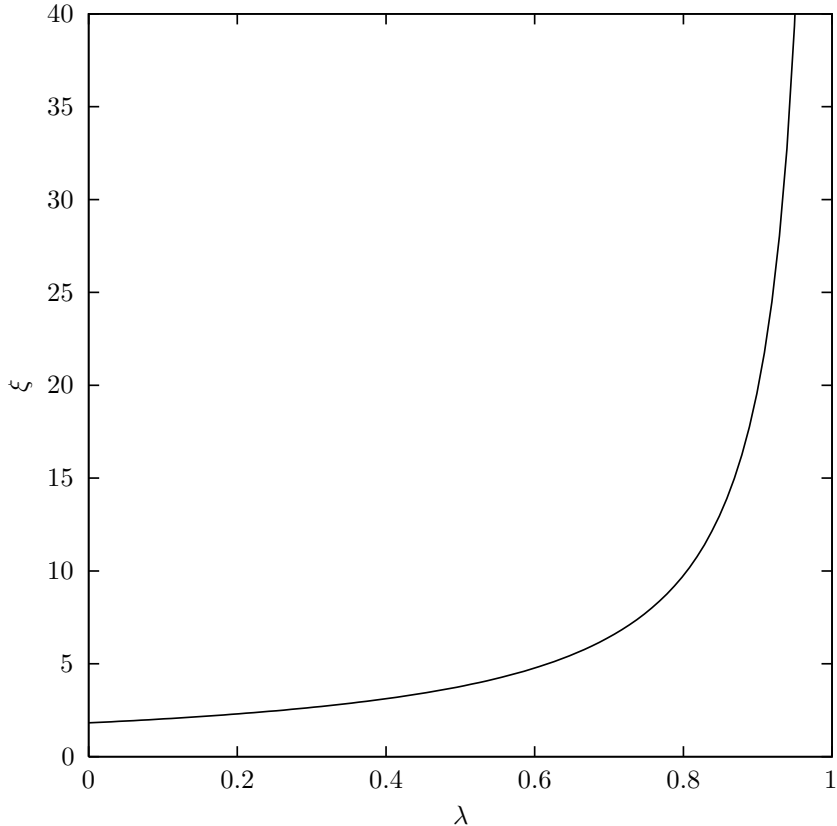}
\end{center}
\caption{\label{fig:onedlengthscale}The size of the bound-state hole pair $\xi$
(in units of the lattice parameter)
on the one-dimensional chain, as a function of the model parameter, $\lambda$.
The binding energy vanishes as $\lambda \rightarrow 1$, with a concomitant divergence
in the real-space size of the bound-pair.}
\end{figure}

The Greens function contains the eigenvalue \emph{and} eigenvector information, so this
technique allows us to determine the `size' of the bound-state pair as a function of binding energy.
Provided that $\delta\neq 0$, the amplitude of the real-space wavefunction when the holes are separated
by $\Rvec_i$ is proportional to
\begin{equation}
\Greens_{i0}(\boundstateenergy) = \frac{1}{N} \sum_{\kvec} 
\frac{\exp[\rmi \, \kvec . \Rvec_i]}{\boundstateenergy + \lambda - \gamma_{\kvec}}\,.
\end{equation}
In the one-dimensional example, this quantity is
\begin{equation}
\Greens_{i0}(\boundstateenergy) \propto
\frac{(-1)^{\vert i \vert}}{2\lambda - \sqrt{\lambda^2 + 3}}
\left [
\frac{1}{3} \left( \lambda + \sqrt{\lambda^2 + 3}  \right)
\right]^{\vert i \vert} ,
\end{equation}
which means that, for $\lambda < 1$, the magnitude of the
real-space wavefunction decays exponentially with a length scale $\xi$ defined by
\begin{equation}
\frac{1}{\xi} = - \ln \left[ 
\frac{1}{3} \left ( 
\lambda + \sqrt{\lambda^2+3}
\right) \right] \, .
\end{equation}
The variation of this length scale with $\lambda$ is plotted in figure \ref{fig:onedlengthscale}.
%
%
\section{Discussion and conclusions}

We have shown that, for a restricted range of $\lambda$, pairs of holes in our strong-coupling Hubbard
model bind together to form the real-space analogue of Cooper pairs.
The background spins are induced into an RVB state by the motion of these holes.
We have not addressed the important issue of the relevance of these results to superconductivity in
\CuOtwo\ planes, and whether or not one would anticipate superconductivity in our model at
finite doping~\cite{longchampion2003}.

It is important to realize that the demands of superconductivity are less stringent than
that of binding holes into pairs. In one and two dimensions any magnitude of net attraction
between two holes guarantees a bound-state, but in three dimensions there is a non-zero
threshold that must be exceeded.
Consequently, we can achieve the bound-state for the linear chain and square superlattice,
but in the cubic superlattice case the bound-state is irrelevant as it only forms for values
of $\lambda$ for which the groundstate is ferromagnetic.

Superconductivity, on the other hand, involves macroscopic doping. When the number of holes
is of the same order as the number of lattice sites, the holes are forced to be near each
other, and then can always make use of an attraction to form a `collective' paired state:
attraction guarantees a pairing instability of a Fermi surface. It is possible that in one
dimension a Luttinger liquid might arise, with power-law pairing correlations and no
long-range order, but inter-layer coupling stabilizes the superconductor.
%
\ack We thank the organizers of SCENM02 for providing the opportunity to present this work.
JDC acknowledges financial support from the University of Birmingham and the EPSRC.

%
\end{document}